\documentclass[sigconf]{acmart}

\usepackage{booktabs} 

\acmConference[CIDR'19]{8th Biennial Conference on Innovative Data Systems Research}{January 2019}{Asilomar, CA, USA.}
\acmYear{2019}
\copyrightyear{2018}

\begin{document}

\title{Modeling Corruption in Eventually-Consistent Graph Databases}

\author{Jim Webber}
\affiliation{
  \institution{Neo4j}
  \city{41-45 Blackfriars Rd \\ London}
  \country{UK}
}
\email{jim.webber@neo4j.com}

\author{Paul Ezhilchelvan}
\orcid{0000-0002-6190-5685}
\affiliation{
  \institution{School of Computing \\ Newcastle University}
  \city{Newcastle Upon Tyne}
  \country{UK}
}
\email{paul.ezhilchelvan@ncl.ac.uk}

\author{Isi Mitrani}
\affiliation{
  \institution{School of Computing \\ Newcastle University}
  \city{Newcastle Upon Tyne}
  \country{UK}
}
\email{isi.mitrani@ncl.ac.uk}

\begin{abstract}
We present a model and analysis of an eventually consistent graph database where loosely cooperating servers accept concurrent updates to a partitioned, distributed graph. The model is high-fidelity and preserves design choices from contemporary graph database management systems. To explore the problem space, we use two common graph topologies as data models for realistic experimentation. The analysis reveals, even assuming completely fault-free hardware and bug-free software, that if it is possible for updates to interfere with one-another, corruption will occur and spread significantly through the graph within the production database lifetime. Using our model, database designers and operators can compute the rate of corruption for their systems and determine whether they are sufficiently dependable for their intended use.
\end{abstract}



\maketitle

\section{Introduction}

Graphs have long been a useful mathematical abstraction and commonplace data structure. More recently modeling data as a graph has become fashionable and graph databases have become an popular technology choice\cite{dbengines}. Using edges and vertices to create high fidelity data models is expressive and straightforward, even for complicated problem domains. Furthermore querying graphs can be made fast and efficient on modern computer hardware.

Today there are many graph databases available under open source and commercial licenses. Each graph DBMS designer makes their own design choices about how to manage graph data. For example, the most popular database the category is Neo4j\cite{neo4j}, an efficient database for graph queries which supports ACID transactions for updates. However, Neo4j is a \textit{replicated} graph database, which imposes practical limits to its scale (on commodity storage) but is designed to be safe even in the presence of faults. Other systems such as JanusGraph \cite{JanusGraph}, TitanDB \cite{TitanDB} \textit{partition} graph data across underlying NoSQL data store instances (often Apache Cassandra \cite{Cassandra}) such that each instance manages a subset of the graph and processes a portion of the workload. Accordingly they should be more scalable, but with weaker consistency guarantees.

Our concern is that databases designed for key/value, column, or document storage may not be sufficiently \textit{dependable}\cite{avizienis} for graph data even if a graph can be encoded into its primitives. Graphs are demanding of their underlying DBMS since they have explicit relationships as well as data values which must be carefully maintained. As we shall see, maintaining structural integrity across replica sets --- what we term \textit{reciprocal consistency} --- in a distributed graph is non-trivial and failure to maintain reciprocal consistency leads to data corruption.

In the remainder of this paper we discuss a popular design pattern for distributed graph databases and show how that pattern is susceptible to corruption under weak isolation (prominent in eventually consistent systems). We further demonstrate how further semantic corruption arises and show how it quickly spreads, based on our prior work \cite{EPEW2018TR} using simulations and (approximate) numerical analyses. We end our discussion with some practical considerations for graph users and those involved in the design of contemporary eventually consistent graph databases.

\section{Implementation of Distributed Graph Databases}

A distributed graph database partitions a graph between a number of loosely cooperating servers. A common design (arising from \cite{JanusGraph} and TitanDB \cite{TitanDB}) is for records to represent vertices containing both data values and an adjacency list containing edge ``pointers'' to other vertices. An edge E joining two vertices A and B results in \textit{reciprocal} entries for E being written in the adjacency lists for both A and B such that a query reading either A or B would be able to reify the edge correctly. When the adjacency lists for vertices are mutually compatible like this, we say they are \textit{reciprocally consistent}.

Reciprocally consistent adjacency lists provide directed edges which can be queried in either direction (from A or B) at the same cost. Traversing edges in both directions is typical for graph workloads where, at its simplest extent, we might ask not only ``who do I follow on Twitter?'' but also ``by whom am I followed?''

Most edge pointers in the adjacency lists refer to local vertices on the same server. The rest will refer to vertices on other remote servers. Minimizing the number of such inter-server edges is the principal objective of good partitioning algorithms \cite{Leopard2016, Firth2017} because remote edges cost more for queries to traverse (network penalty) and, as we shall describe, are also costly (in time) to store safely.

To ensure edge records are reciprocally consistent, a graph database could update distributed records through protocols like Paxos \cite{lamport1998}, RAFT\cite{Raft2014}, or 2-Phase Commit \cite{Gray1978, Lampson1981}, and provide isolation with locks or MVCC. However, the scalability of the database will be impacted, which is hard to reconcile with the demands of shared-nothing scale-out use-cases. Instead, most contemporary distributed graph databases (e.g. Titan, JanusGraph) use an existing eventually consistent database (like Apache Cassandra) to store data. They adapt the underlying database with a programmatic API or query language expressed in terms of edges and vertices along with some gluecode to bind that interface to the underlying DBMS.

Superficially this appears to be a good choice: the user has the modeling convenience of graphs with the operational characteristics from the underlying database. However, eventual consistency semantics pertain to the convergence of single record (document, column, or key-value pair): given sufficient time (and inactivity) the contents of a record become consistent across all its replicas\cite{Bailis2013}. There is no such convergence guarantee between the two reciprocal adjacency list records that constitute a distributed edge.  

\section{Dependability Failings of Eventually Consistent Graph Databases}

A distributed edge relies on two reciprocally consistent adjacency lists stored in records in two different replica sets. We assume that the replica sets themselves always converge and that practically there are mechanisms like read/write quorums to prompt intra-replica convergence.

\begin{figure}
  \includegraphics[width=\linewidth]{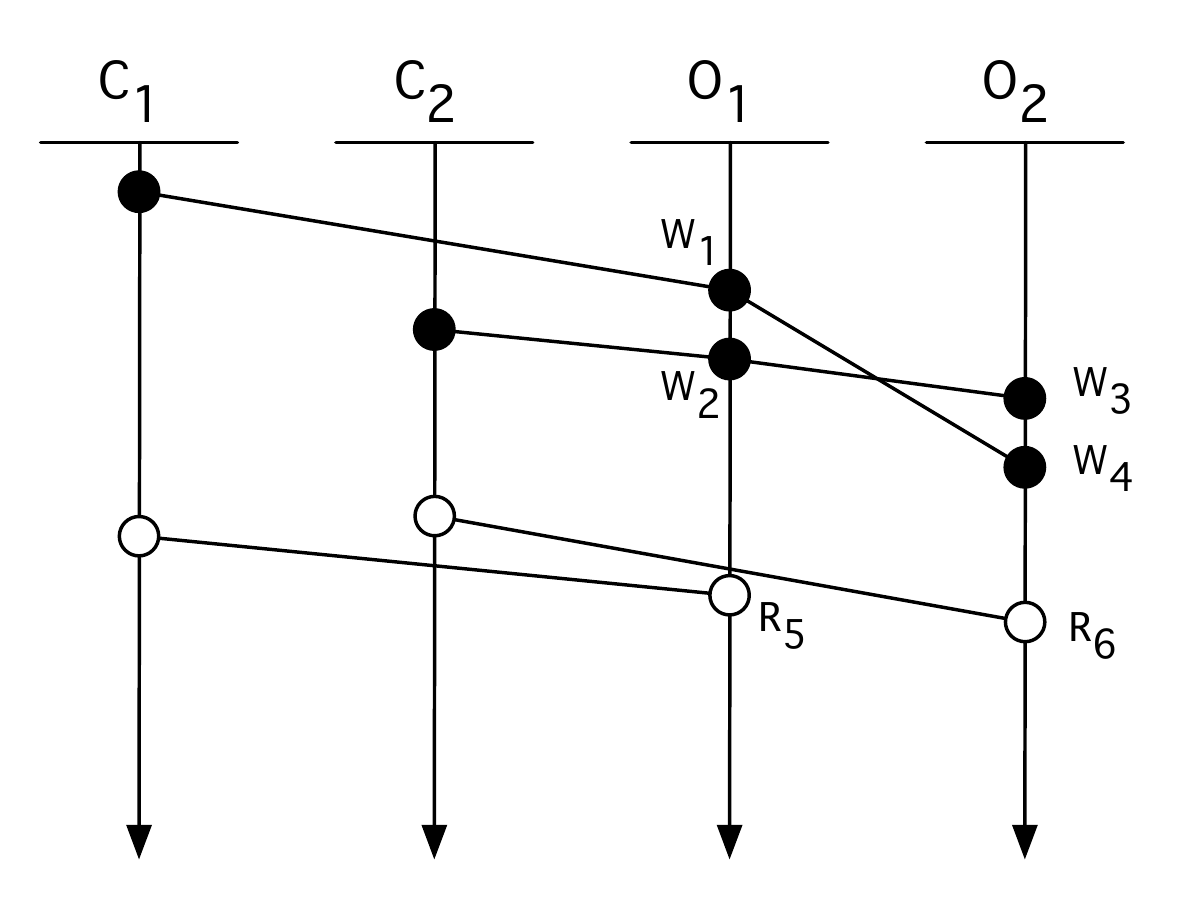}
  \caption{Overlapping updates on two reciprocal vertex records.}
  \label{fig:f1}
\end{figure}

However, eventually consistent databases provide no guarantees of reciprocal consistency \textit{between} replica sets. Furthermore there is no general mechanism for ensuring \textit{isolation} between concurrent updates on popular eventually consistent databases. With stricter consistency models like ACID we can reason about numerous isolation levels to suit the workload that can prevent updates from interfering with one-another. Under eventual consistency semantics, updates can be interleaved across replica sets which can have unhelpful side-effects. Sometimes we may be fortunate that interleaved updates commute, and sometimes --- with a non-zero probability --- we are not so fortunate and reciprocal consistency is lost. We show this phenomenon in Figure \ref{fig:f1}.

Consider two records O1 and O2 which represent connected vertices in a graph. They are concurrently accessed by clients C1 and C2 and each of O1, O2, C1 and C2 execute on distinct servers. To preserve correctness, each server should process updates in the same order (unless the operations are somehow known in advance to be commutative). That is, C1 should write W1 to O1 and W4 to O2 before C2 writes W2 to O1 and W3 to O2 maintaining reciprocal consistency\footnote{Equally, C2's writes could precede C1's and, assuming that there is no causal link between them, the writes would still preserve correctness as we see in protocols like \citet{WARP2015}}.

Unfortunately it is possible for writes to interleave in this scenario because there is no isolation between them: C1 writes W1 to O1, followed by C2 writing W2 to O1, followed by C2 writing W3 to O2, and finally C1 writing W4 to O2. Writes W3 and W4 arrive in the opposite order with respect to W1 and W2.

Any non-commutative overlapping updates will lead to divergence (a polite term for corruption). We consider this to be \textit{reciprocally inconsistent} or \textit{mechanical corruption}: the machinery has, by design, partially corrupted data\footnote{It may be possible to detect mechanical corruption by comparing pairs of records across servers, but there is no guarantee that the corruption can be correctly repaired back to the linearizable truth since the (transactional) context of any writes is lost and cannot generally be reconstructed post-hoc. We may call this the ``Back to the Future 2'' problem on the basis that repairing history can actually worsen the present state of the world.}.

\begin{figure}
  \includegraphics[width=\linewidth]{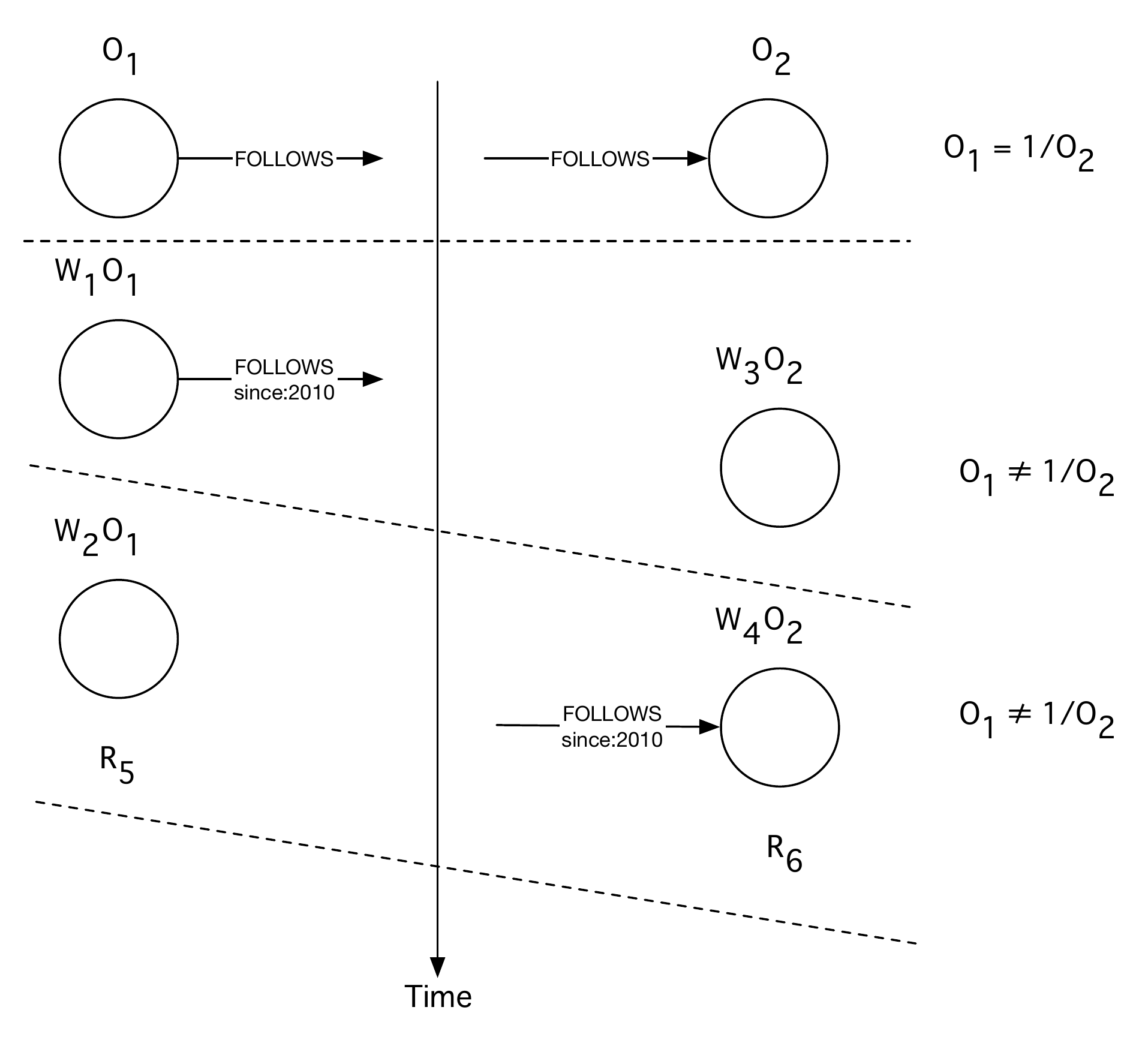}
  \caption{Reciprocally inconsistent vertex records due to interleaved updates.}
  \label{fig:f2}
\end{figure}

If we read based on O1 we see a different graph than if we read based on O2. In Figure \ref{fig:f2} an update edge write and a delete edge write arrive out of order leaving the graph in a superposition of (conflicted) states. Should we pick the ``wrong'' edge record (with respect to linearizable truth) on which to base subsequent writes then we introduce semantic garbage into the graph. That garbage itself can be used by subsequent queries, spreading semantic corruption through the graph. Conversely, should a query happen to select the ``correct'' edge then no semantic corruption will be introduced, but the subsequent write may also be subject to interference with other updates.

Unfortunately, anti-entropy mechanisms like quorums don\textquotesingle t help. If O1 and O2 are sets of replicated objects \{$O1_1$, $O1_2$, ... $O1_n$\} and \{$O2_1$, $O2_2$, ...$O2_n$\} then the anti-entropy mechanisms will help individual replica sets to converge to the same state but they do not help inter-replica set convergence for a consistent reciprocal state. It can be stated using the temporal operator $\diamond$ (for eventually) as:

\begin{itemize}
  \item O1 has eventually converged: $ \diamond  \nexists i,j : O1_i \neq O1_j$, and 
  \item O2 has eventually converged: $ \diamond  \nexists i,j : O2_i \neq O2_j$, and
  \item O1 and O2 are not reciprocally compatible: $ \diamond \forall i,j : O1_i \neq 1/O2_j$ 
\end{itemize}

It might be the case that writes interfere very rarely and so the rate of corruption is low enough not to be a problem in practice. Our next step is to determine the rate of corruption and satisfy ourselves whether eventually consistent graph databases are sufficiently dependable for any given use-case.

\section{Modeling}

To predict what will happen to an eventually consistent graph database under load, we designed a probabilistic model \footnote{We chose not to perform an empirical evaluation of existing systems because such an evaluation would have been impractical (need to compare database state at end of experiment with the linearizable truth), slow (real time) and expensive (requiring many hours of storage and compute time).} that allows us to examine the behavior of the system. The model and its analysis are detailed in  \cite{EPEW2018TR}. We explain here how our model construction captures the operational aspects of an eventually consistent database, and also present and rationalize the approximations taken in analytically evaluating the system behavior. Loss of accuracy due to these approximations is shown to be less than 1\% in \cite{EPEW2018TR}.

For simplicity, the model assumes that the hardware never crashes and the software is bug free \footnote{The shortcomings of imperfect hardware and software can be dealt with through redundancy in practice, but that is essentially orthogonal to the model.}. Though these assumptions present the system in a more favorable light than in practice, they help us to focus on examining corruption behavior that can be solely attributed to eventual consistency.

\begin{figure}
  \includegraphics[width=\linewidth]{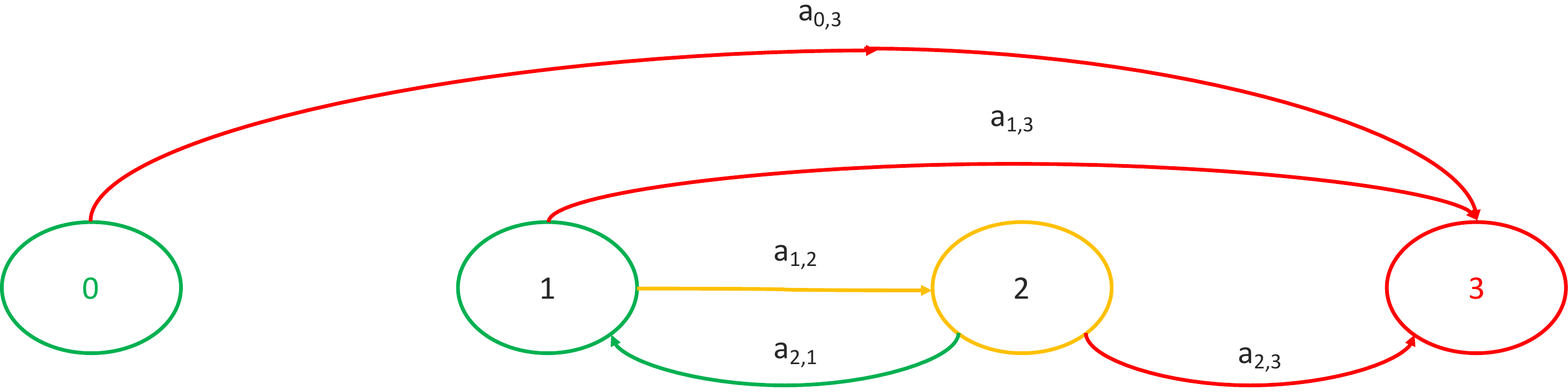}
  \caption{Transitions between good, partially corrupt, and semantically corrupt record states.}
  \label{fig:f3}
\end{figure}

Figure \ref{fig:f3} shows the permitted state transitions for records. States 0 and 1 are good initial states for local and distributed edges respectively. State 2 is a reciprocally inconsistent distributed edge arising from overlapping (non-isolated) updates. State 3 represents semantic corruption of edges - be local or distributed. It is also the `absorbing' state because no general graph repair algorithm exists to restore the database from state 3  back to the linearizable truth. 

Of interest is the average first passage time $U_\gamma$ for the ratio of the number of records in state 3 to the total number, to reach or exceed a given fraction $\gamma$ (say, 10\%). Informally, an initially corruption-free database is regarded to have become too corrupt to be useful in practice after an operational period of $U_\gamma$.

Labels $a_{i,j}$ in Figure \ref{fig:f3} denote the average rate at which records in state $i$ enter state $j$, $i \neq j$ and $0 \leq i, j \leq 3 $, under a given work load. Since we assume, again for simplicity, that a distributed graph is not repartitioned, transition between states 0 and 1 cannot occur. A reciprocally consistent distributed edge (in state 1) becomes reciprocally inconsistent (and enters state 2) when it experiences uncoordinated conflicting updates by concurrent queries as explained earlier; that is, before a query could complete its distributed write at the remote end, another query starts its update at that end and leaves the edge in state 2. Since local writes are (presumed) instantaneous, transition from state 0 to state 2 cannot occur.

If a query reads the ``wrong'' end of a state 2 edge and then updates edges, all updated edges, local or distributed, enter state 3 (if they are not already in that state). Thus the preconditions for the first transition to state 3 are the existence of a state 2 edge and its wrong edge being used (with probability $\frac{1}{2}$) for subsequent updates\footnote{So, graph repartitioning need not repair a state 2 edge: if it makes a state 2 edge into a local edge and chooses the wrong end as the new value for that edge, it would only turn a state 2 edge into a semantically-corrupt local edge in state 3.}. Once state 3 edges come into being, they also contribute to propagation of semantic corruption like state 2 edges, but with a crucial difference: state 3 edges have no ``correct'' or ``incorrect'' sides since corruption is semantic - reading either side prior to updates causes all the updated edges to enter state 3.

While reading corrupt records prior to writing good edges leads to further corruption, reading \emph{only} good records before overwriting reciprocally inconsistent edge records results in correction, provided that the overwriting itself is conflict-free. When a query has read only state 0, state 1, or the correct side of state 2 edges, and then overwrites a state 2 edge in a conflict-free manner, it corrects that state 2 edge back to state 1. That is, it is possible for the mechanical corruption of a state 2 edge to be corrected, provided that the all prior reads are on clean records \emph{and} the distributed write is conflict free; however, if the latter fails to hold, the overwritten edge continues to remain in state 2 albeit with different, reciprocally-inconsistent values at the end records.


In summary, distributed edges are subject to two competing trends: given that all reads prior to an update are on clean records, a conflicted write would transfer them to, or keep them in, state 2 and a conflict-free one, on the other hand, would transfer them to, or keep them in, state 1. Further, if any of the earlier reads by a query has involved a corrupt record, any edge in state 0, 1 or 2 which  gets updated subsequently, enters state 3 permanently.

\subsection{Database Corruption} \label{subsec:DBCorruption}

Since the emergence of state 2 distributed edges cannot be ruled out when strict isolation is not guaranteed, the graph database must become irredeemably corrupt for all values of $\gamma$ \textit{eventually}, i.e., for some finite values of $U_\gamma$. Of interest, therefore, is: how large or small is the value of $U_\gamma$ for a given value of $\gamma$? The answer depends on several parameters characterizing three systemic aspects:

\emph{Size and topology of graph database}. Size is expressed by the total number of edges ($N$), and the fraction ($f$) of edges distributed across servers.  
As  explained later, we consider two distinct edge access patterns or \emph{topologies}: in \emph{Complete} topology, all edges are equally likely to be accessed by a query, while, in a \emph{Scale-free} topology, edges have different access probabilities and those that get accessed more frequently tend to be smaller in number. 
      
      %

  \emph{Work load} is usually measured as transactions per second or TPS. Significant for $U_\gamma$, however, are: the fraction of this load that writes after reads and the number of reads that precede a write.
  
\emph{Distributed Write Delays}. The smaller they are, the less likely is that writes by queries overlap. Thus, the average value and the distribution of these delays are also influential.

Our objective is to create an easy-to-use tool to compute $U_\gamma$ and we considered multiple approaches. First, we simulated the system and a single run took too long to complete even when $N$ is in millions.  
So, we attempted exact analytical evaluation of the model which appeared infeasible. 
With some justifiable approximations, an analytical solution was found. It involved solving a set of non-linear simultaneous equations using fixed point iterations. Brouwer's theorem \cite{bro} guarantees that a solution exist. 

We assessed the effect of our approximations by using simulations. The loss of accuracy was found to be negligible ($<$ 1\%). Thus, the approximate analysis meets our objective. 
Next subsection sketches the important aspects of our analytical approach and explains the approximations. For brevity, we consider only the complete topology and scale-free graphs are dealt with in \cite{EPEW2018TR}.

\subsection{Analytical Approach and Approximations} \label{subsec:evaluationApprox}

Denote by $n_{i}(t)$ the number of state $i$ edges at time $t$. The vector $\mathbf{n}(t) =[n_{0}(t)$,$n_{1}(t)$,$n_{2}(t)$,$n_{3}(t)]$ defines the state of database at time $t$. At all times, the elements of $\mathbf{n}(t)$ add up to the total number of edges $N$. The database is initially clean; so, $\mathbf{n}(0)=[(1-f)N,fN, 0 , 0]$. $U_\gamma$ is the smallest $t$ when $n_{3}(t)\geq \gamma N $.

We assume that the queries arrive in a Poisson stream and 10\% of them update edges. Thus, the queries of interest have a Poisson arrival rate of $\lambda = $ 10\% of TPS. Each of these queries is (pessimistically) assumed to write exactly once after having performed 2 or more reads whose number is geometrically distributed with parameter $r$.

A read operation performed by a query at time $t$ would be on a clean record with probability $\alpha = \frac{1}{N} [ n_{0}(t)+n_{1}(t)+\frac{1}{2} n_{2}(t)]$. It is the probability that the read operation picks up a state 0 edge
(probability $n_{0}(t)/ N $) or a state 1 edge (probability $n_{1}(t)$/$N $) or the correct side of
a state 2 edge (probability ($1/2$) $\times$ $n_{2}(t)/ N $).

The probability, $\beta$, that all read operations by a query arriving at time $t$ would be clean, can be evaluated (using properties of geometric distribution starting at 2) as: $\beta = \alpha^{2} r / [1 - \alpha (1-r)]$.

Consider now the probability, $q$, that a query, say, $Q$ arriving at time $t$ and taking a time $d$ to complete a distributed write, will be involved in a conflict with another query $Q'$. 

When $d$ is exponentially distributed with mean $\delta$, two exponential processes are in progress: (i) $Q$ completing its distributed write at rate $1/ \delta$ and (ii) arrival of $Q'$ at rate $\lambda (1/2N)$. The rate for (ii) is obtained by conditioning the arrival rate $\lambda$ on $Q'$ choosing one specific edge out of $N$ edges and on choosing one specific end of that edge (which occur with probabilities $1/N$ and $1/2$, respectively).

The probability that (ii) occurs before (i), is the ratio of rate of (ii) to the sum of rates of (i) and (ii). Thus,  $ q = (\lambda \delta/ [2N+\lambda \delta])$. 

\textbf{Transition rates} $a_{i,j}$ of Figure \ref{fig:f3} can now be expressed by
multiplying the arrival rate $\lambda$ by the probabilities of all events required for an edge to transit from state $i$ to state $j$.
For example, $a_{i,3}$, $i = 0, 1, 2$, would be $\lambda \times (n_{i}(t)/ N) \times (1- \beta)$; here,
the probability that a query updates a state-$i$ edge (out of $N$ edges) and at least one of its earlier reads be on a corrupt record is $(n_{i}(t)/ N)  \times (1- \beta)$.

In general, $a_{i,j}$ can be expressed as: $g_{i,j} \times n_{i}(t)$,
where $g_{i,j}$ is some function of $N$, $\beta$ and $q$,
and $n_{i}(t)$ is the number of edges in the `from' state of the transition.
Since a transition to state 3 is independent of whether the update is conflicted or not,
$g_{i,3}$ is not a function of $q$;
also, $g_{0,3} = g_{1,3} = g_{2,3} = g_{*,3}$ (say).
Transitions between states 1 and 2, on the other hand,
require a conflicted or conflict-free write,
and $q$ is a factor in $g_{1,2}$ and $g_{2,1}$.

\textbf{Fluid approximation}. Instead of describing the system state by integer-valued
functions,
it is convenient to use fluids
being in different states or `buckets'.
So, $n_{i}(t)$ now becomes a
real-valued function indicating the amount of fluid present
at time $t$ in a bucket called state $i$
($i=0,1,2,3$).
Fluids flow continuously out of, and into buckets, at transition rates $a_{i,j}$.

Define $n_{i}^\prime(t) = \mathrm{d}n_{i}(t)/ \mathrm{d}t$ as the rate
at which $n_{i}(t)$ \textit{increases} at $t$.
It is the sum of all inflow rates (into bucket $i$)
\emph{minus} that of outflow rates. Thus, $n_{0}^\prime(t) = - a_{0,3} = - g_{0,3}n_{0}(t)$, $n_{1}^\prime(t) = g_{2,1}n_{2}(t)- (g_{1,3}+ g_{1,2})n_{1}(t)$, 
$n_{2}^\prime(t) = g_{1,2}n_{1}(t)- (g_{2,3}+ g_{2,1})n_{2}(t)$, and $n_{3}^\prime(t) = g_{*,3}[n_{0}(t)+ n_{1}(t) + n_{2}(t)$.
Since $n_{3}^\prime(t) > 0$, a finite $U_\gamma$ must exist.

\textbf{An approximation for tractability}.
The expression for $n_{0}^\prime(t)$ contains only $n_{0}(t)$
and not any other $n_{j}(t)$, $j \neq 0$;
on the other hand, those for $n_{i}^\prime(t)$, $i = 1, 2, 3$
contain some $n_{j}(t)$, $j \neq i$.
This inter-coupling means that the
differential equations may not be solved using known methods.
Complexity is also exacerbated by $\alpha$ and $\beta$ being
time-dependent and also appearing in a non-linear form: a state 1 edge entering state 2, for example, requires \emph{both} the conflicting queries to have read only clean records and thus $g_{1,2}$ is a function of $\beta^2$.

So, we take the approximation of replacing any `inconvenient'
$n_{j}(t)$ by its average $\bar{n}_{j}$ during [$0$,$U_\gamma$]:
$\bar{n}_{j} = \frac{1}{U_\gamma}\int_0^{U_\gamma} n_{j}(t)dt$.


For example, $n_{1}^\prime(t)$ can now be written as:
$n_{1}^\prime(t) = g_{2,1}\bar{n}_{2}- (g_{1,3}+ g_{1,2})n_{1}(t)$
where $\bar{n}_{2}$ is independent of $t$.
Similarly, we also make $\alpha$, and hence $\beta$,
independent of time; e.g.,

$\alpha = \frac{1}{N} [ \bar{n}_{0} + \bar{n}_{1} +\frac{1}{2} \bar{n}_{2}]$.


Solving the differential equations is now possible and leads to
four, non-linear simultaneous equations, one for each $\bar{n}_{i}, i = 0, 1, 2$ and one
for $U_\gamma$.
The latter are solved using 
\textbf{fixed point iterations} as explained below.

Start with some initial estimates for $\bar{n}_{i}$; call
them $\bar{n}_{i}^{(0)}$. 
Using the new expression for $\alpha$ above,
get an initial
estimate for $\alpha$ and hence for $\beta$; call those
$\alpha^{(0)}$ and $\beta^{(0)}$. Then, compute an
initial estimate for $U_\gamma$, called $U_\gamma^{(0)}$. In step $j, j > 1$ of this procedure, the values $\bar{n}_{i}^{(j-1)}$,
$\beta^{(j-1)}$ and $U_\gamma^{(j-1)}$ are used to compute $\alpha^{(j)}$,
$\beta^{(j)}$, $\bar{n}_{i}^{(j)}$ and $U_\gamma^{(j)}$. The process
terminates when results of two consecutive iterations are
sufficiently close. 
A solution must exist by Brouwer's theorem \cite{bro} since $\bar{n}_{i}^{(j)}$ are
bounded and the mapping $\bar{n}_{i} \rightarrow \bar{n}_{i}$ as defined by the simultaneous equations is continuous.



\section{Evaluation}

Our model assumes that the system processes many reads-followed-by-write graph queries concurrently, processing both local and distributed edges. We chose two graphs with \textit{Scale-Free}\cite{scalefree} and \textit{Complete}\cite{complete} topologies to experiment upon. Both graphs were large enough to make processing non-trivial:

\begin{enumerate}
  \item A large Scale-Free graph, such as a social network, human brain, or road network
    \begin{itemize}
      \item Approximately 7.7 billion local edges
      \item Approximately 3.3 billion distributed edges (in proportion to good graph partitioning algorithms)
      \item Approximately 1 billion nodes
    \end{itemize}
  \item A large Complete graph where every vertex is connected to every other as a baseline. 
    \begin{itemize}
    \item Approximately 7 billion local edges
    \item Approximately 3 billion distributed edges (in proportion to good graph partitioning algorithms)
    \item Approximately 1 billion nodes
  \end{itemize}
\end{enumerate}

In the Scale-Free graph, edges are classified by popularity - how frequently they are accessed by queries. There is a small set of very popular edges (e.g. a link between two famous people, connections between brain centers, or roads between large cities) through several grades down to a very large number of unpopular edges (e.g. a link between unpopular computer scientists).

We consider 7 categories of popularity, $ j = 0, 1, \ldots, 6$, with $j=0$ and $j=6$ being the most and the least popular respectively. Category $j$ has $N_j = 10^{4+j}$ edges and access probability  $p_j = 1/2^{j+1}$ rounded to 2 decimal points such that $\sum_{j=0}^{6} p_j = 1.0$. Our aim is to measure how topology affects the spread of corruption through the data, as load would be biased towards popular edges.

By also considering the Complete graph, we negate the graph's topology as an influence for directing corruption. Since every node is connected to every other, the effect is a random access to the graph with all nodes being equally likely to be encountered next during a query execution. 

The Poisson arrival rate $\lambda$ of queries that write once after a few reads is varied from 1,000 to 10,000 per second which is equivalent to varying TPS from 10,000 to 100,000 if $\lambda = 10\%$ of TPS. Number of reads prior to a write is at least 2 and geometrically distributed with parameter $r=0.4$, amounting to an \textit{average} of 2.5 reads before write. We believe this is proportionate for non-trivial graph queries. 

The time to write a distributed edge is chosen to be exponentially distributed with mean $\delta =$ 5ms to reflect network and disk access delays. So, the probability $q$ of any query conflicting with another during a distributed write is approximately 1/50,000 when $\lambda = 1000$ (or TPS = 10,000). Finally, the database is regarded to be corrupt when $\gamma = 10\%$ of edges reach state 3 (semantic corruption)\footnote{Another, more nuanced, option would be to measure the proportion of queries that encounter corrupted records. However we felt that ratio of corrupted records is a more straightforward metric for operators and in either case does not fundamentally change the outcome.}.

\section{Results}


\begin{figure}
  \includegraphics[width=\linewidth]{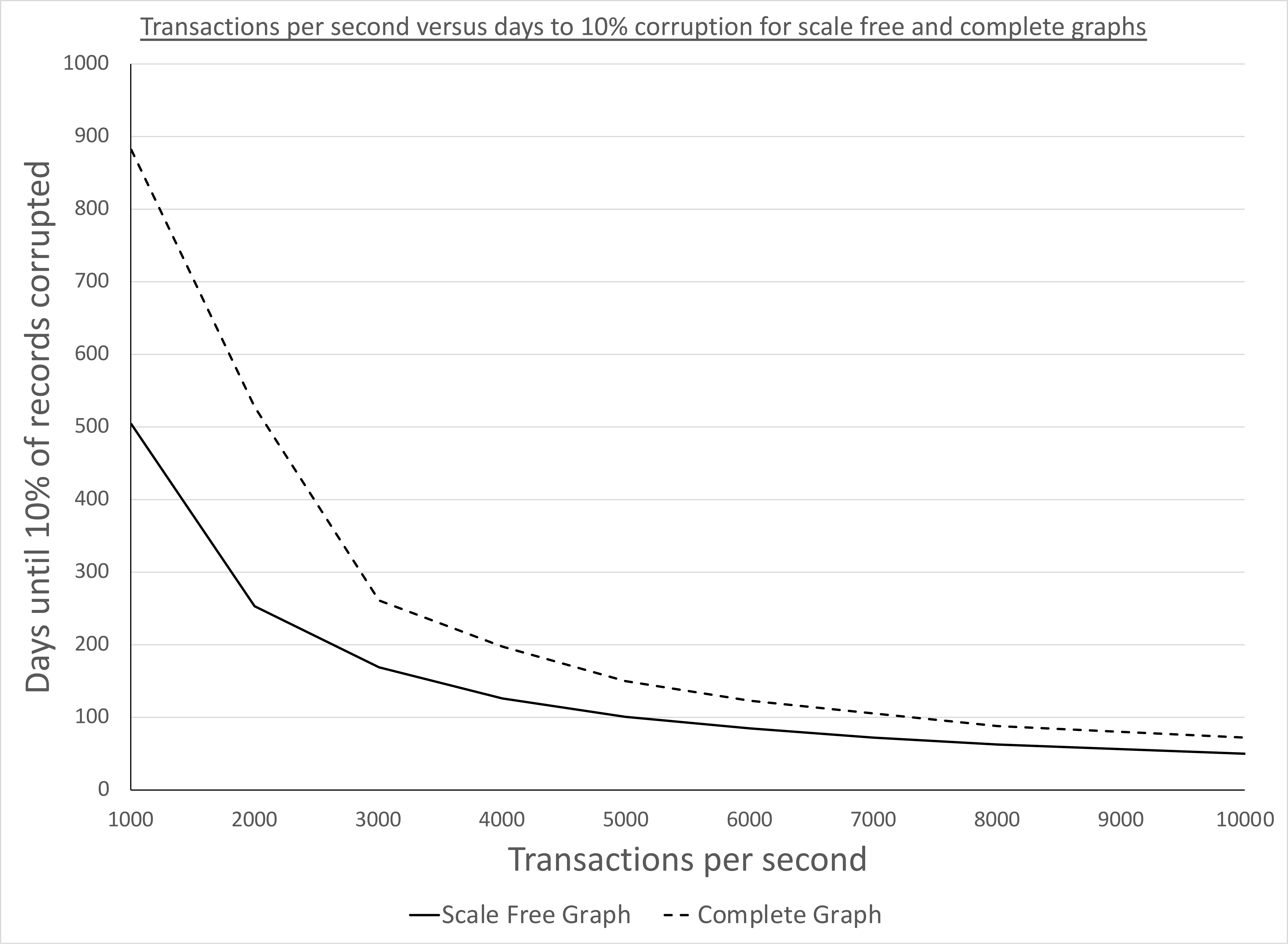}
  \caption{Time in months for 10\% of semantic corruption ($U_{0.1}$) versus arrival rate $\lambda$ of reads-before-write queries.}
  \label{fig:f4}
\end{figure}

It is significant from Figure \ref{fig:f4} that corruption occurs in \textit{months} even for modest arrival rates, e.g., $2000 \leq \lambda \leq 3000$. This means that we can expect significant data corruption within the lifetime of a production database. Most eventually consistent graph databases claim high scalability so it is plausible that arrival rates could be much higher than our experiment and corruption would happen more quickly. Equally smaller graphs under the same load will be corrupted more quickly (since we need fewer corrupted records to meet our 10\% threshold). Moreover graph repair \textemdash even if it could be effected safely \textemdash or reducing the time cost of writes will delay but cannot indefinitely postpone corruption.

Interestingly, graph topology did not have as profound an impact on rate of corruption as we had expected. The Scale-Free graph focuses load around popular edges. As such those edges corrupt relatively quickly. However, 
it takes quite some time to spread that corruption to the much larger number of unpopular edges that are nevertheless accessed far less frequently (e.g., $N_6 = 10^{6} \times N_0$, $p_6 = p_0 / 2^6 $) and tip the database into a 10\% corrupt state.

The Complete graph topology does not favor writes in any specific neighborhood and so it takes more time for conflicted distributed writes to occur and generate the initial seeds of state 2 corruption. But once the latter come into existence, subsequent semantic corruption tracks closely to the Scale-Free graph case. 

We believe that the dominating factor is that graph transactions involve reading multiple records (traversing the graph) before writing records back into the system. In both graphs, queries read 2.5 edges on average (geometric distribution)  before writing. ``Trawling'' the graph like this lets a query read a state 3 or the wrong end of a state 2 edge and then compute a semantically incorrect value to write. Where the Scale-Free graph tends to write back to the popular (and likely already corrupt) records, the Complete graph topology typically writes to an uncorrupted part of the graph chosen at random. Either way, over time the graph topology offers only a minor defense against the spread of corruption.

Finally we observe that even if we were to be more forgiving with some of the model's parameters (fewer overlapping updates, fewer distributed edges), corruption still happens within the lifetime of a production database. This is a cause for concern as it suggests that there are existing production systems that are likely to be corrupting data without their operators being aware.

\section{Practical Considerations} \label{Practical Considerations}

Anecdotally, the most popular eventually consistent database used as the underlay for graph workloads is Apache Cassandra\cite{Cassandra}. While Cassandra is well known as a scalable eventually consistent column store, its designers have also given thought to those situations where eventual consistency is unsuited. In particular, it provides \textit{batched operations}\cite{batchedops} which is an optimistic mechanism to support atomic multi-key units of work and \textit{lightweight transactions}\cite{lwt} which uses Paxos \cite{lamport1998} for conditionally guarded atomic compare and swap (CAS) operations.

Unfortunately both approaches have limitations. Lightweight transactions are limited to a single partition meaning that we can only atomically compare-and-swap on a subset of keys from the database, while the graph's records typically span multiple partitions (for scale). Batched operations \textit{can} span multiple partitions, but in doing so they lose isolation under concurrent load rendering them unsafe for graph writes because they permit overlapping updates. As such they too are only \textit{safe} within a single partition.

For typical Cassandra use-cases, lightweight transactions and batched operations are useful idioms. After all, Cassandra excels for those storage problems which are highly parallel/uncontended and in which the user controls which records (keys) are accessed.

However in a graph query, accessing records is not completely under end user control. A query can traverse far and wide in the graph accessing many records from multiple partitions. Under concurrent load we need to protect queries from mutual interference which is something neither lightweight transactions nor batched operations can safely support.

Whether a specific application can risk weaker consistency from the graph database is not for us to say outright. But we would advise that the users compute their likely time to corruption using our model \cite{EPEW2018TR} so they can determine whether the tendency towards data corruption is problematic for them.

\section{Conclusions}

Given an eventually consistent graph database with weak isolation, the possibility for out of order updates exists. Our numerical analysis shows that this leads to data corruption. Worryingly, corruption spreads quickly enough to be of genuine concern for production database systems.

The model and analytical tools we built can help implementers of eventually consistent distributed graph databases to reason about rates of data corruption given a data volume and workload. In the short term, implementers can parameterize our model to help guide their designs and end users can model their workloads to understand the impacts of the any resulting corruption.

In the longer term we believe that stronger models for graph consistency are required for scalable graph databases. Recent work has shown how to minimize or federate expensive coordination paths in transactional systems\cite{Bailis2015}\cite{WARP2015}. The design and implementation of a distributed graph database based on those ideas is underway in our lab and we look forward to presenting numerical analysis of its dependability characteristics in due course.

\begin{acks}
We are grateful for excellent technical feedback from Hugo Firth, Andrew Jefferson, and Andrew Kerr.
\end{acks}

\bibliographystyle{ACM-Reference-Format}
\bibliography{EC-bibliography}

\end{document}